\documentclass[journal]{IEEEtran}

\ifCLASSINFOpdf
\else
   \usepackage[dvips]{graphicx}
\fi
\usepackage{url}

\usepackage{graphicx}
\usepackage{amsmath}

\begin{document}

\title{Diffuse arrays that autocorrelate and project as delta-like points}

\author{Imants~D.~Svalbe, David~M.~Paganin and Timothy~C.~Petersen
\thanks{Submitted August 2021}
\thanks{Imants~D.~Svalbe is with the School of Physics and Astronomy, Monash University, Victoria, 3800, Australia (e-mail: imants.svalbe@monash.edu)}
\thanks{David~M.~Paganin is with the School of Physics and Astronomy, Monash University, Victoria, 3800, Australia (e-mail: david.paganin@monash.edu)}
\thanks{Timothy~C.~Petersen is with the Monash Centre for Electron Microscopy, Clayton, Victoria 3800, Australia (e-mail: timothy.petersen@monash.edu).}}

\markboth{arXiv Preprint, Nov 2021}
{Shell \MakeLowercase{\textit{et al.}}: Bare Demo of IEEEtran.cls for IEEE Journals}
\maketitle

\begin{abstract}
Diffuse two-dimensional integer-valued arrays are demonstrated that have delta-like aperiodic autocorrelation and, simultaneously, the array sums form delta-like projections along several directions. The delta-projected views show a single sharp spike at the central ray. When such arrays are embedded in larger blocks of two-dimensional data, their location can be fixed precisely via the fast and simple intersection of the back-projected central rays along two or more directions. This mechanism complements localization of the same array from its delta-like autocorrelation, which, although more robust, is slower and more complex to compute.  
\end{abstract}


\IEEEpeerreviewmaketitle
\section{Introduction}
\begin{figure}[htb]
\centering
\includegraphics[width=1.0\columnwidth]{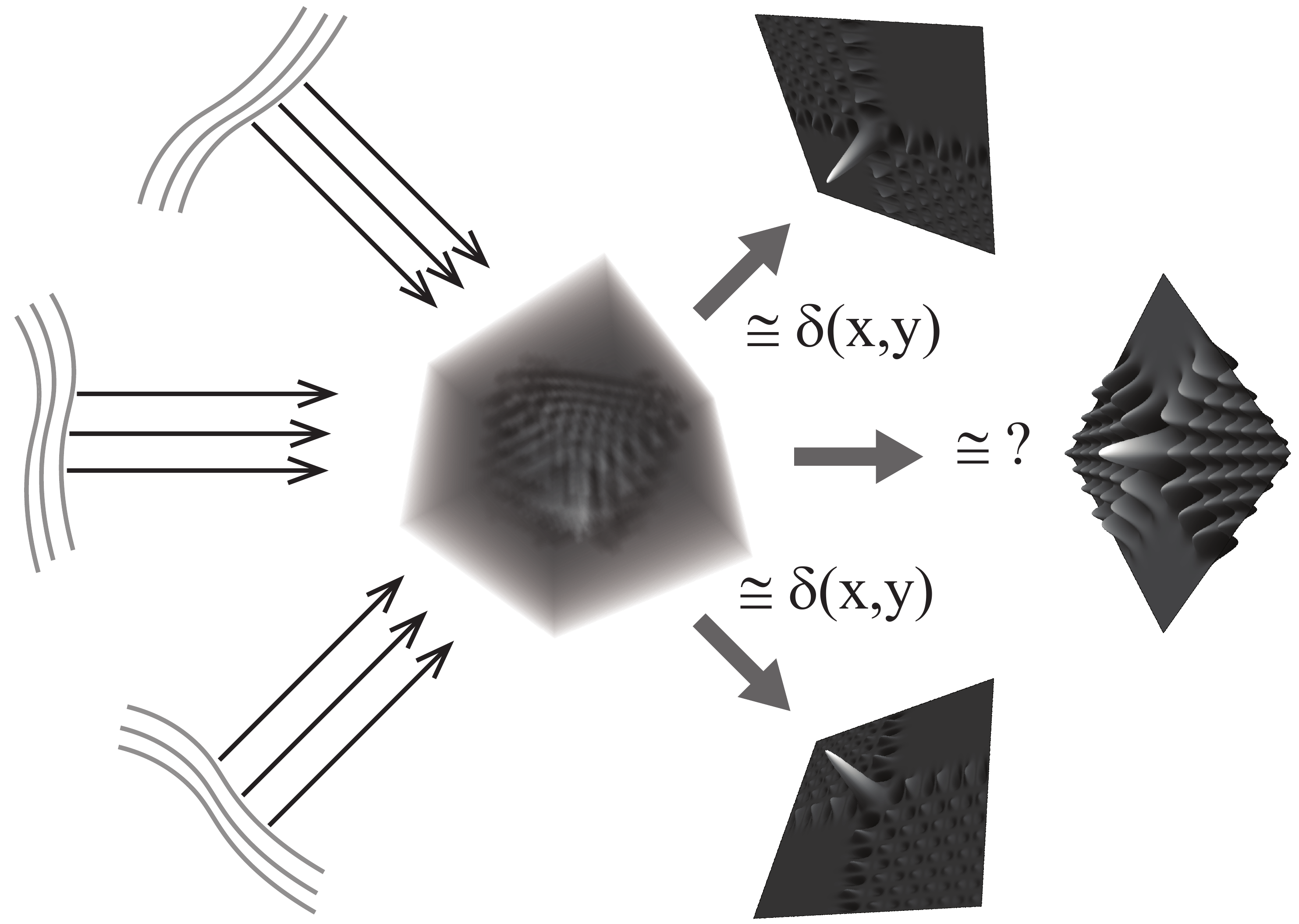} \caption{Delta-like projections arising from a diffuse multi-dimensional function. Spectral flatness enforces additional delta-like correlation.  There is a question as to how many directions can simultaneously project as approximated delta functions, or how these two desired properties of the higher dimensional array can be constructed without significantly concentrating the value of any given voxel.}
\label{fig:Fig1}
\end{figure}

\IEEEPARstart{A}{discrete} or continuum delta function exhibits a perfectly sharp autocorrelation and presents as an ideal sharp point when projected at any view angle.  The former property can be realized for diffuse probes to spread dose while maintaining sharp imaging conditions \cite{SvalbeTCI2020}. This work demonstrates the construction of diffuse discrete integer-valued arrays that emulate \emph{both} of these properties. Two-dimensional delta-correlated arrays have been shown \cite{SvalbeTCI2020} to project as delta functions of the form $[-1,0,\cdots,0,P,0,\cdots,0,-1]$, with peak $P$, along one specific direction.  We show here that this combined correlative and projective mimicry of the delta function can be achieved over several directions.  Fig.~\ref{fig:Fig1} provides an overview of these ideas, whereby a high dimensional array with flat a Fourier spectrum possesses an autocorrelation and lower dimensional projections that are delta-like.  %

This Letter is structured to first review the delta-like correlation properties of Huffman sequences, then describe the formation of arrays with zero sum projections, followed by example arrays, constructed from higher order correlations of canonical Huffman sequences, that have multiple delta-like projections. We conclude by comparing discrete and `analogue' Radon projections of these arrays. 
\subsection{Huffman sequences and arrays}
Huffman sequences \cite{Huffman1962} have optimal aperiodic autocorrelation. A real integer Huffman example, from \cite{HuntAckroyd1980}, is $H_7 = [1, 2, 2, 0, -2, 2, -1]$, with a delta-like aperiodic autocorrelation $H_7\star H_7$ of peak value $18$ and zero values for all off-peak shifts, except the unavoidable end values (-1) that are as small as possible. 

Integer (or real scaled) `canonical' Huffman sequences, $H_N$, of arbitrary length $N = 4n-1$ for positive integers $n$, can be built using recursive sequences \cite{HuntAckroyd1980} or in Fibonacci form \cite{SvalbeTCI2020, svalbeHuffProofs}. Other $H_N$ can be derived from the Fourier spectra of Fibonacci forms, or by direct solution of Diophantine equations expressing the canonical condition \cite{svalbeHuffProofs, HuffmanExtensions2021}.  Extensions to n-dimensions ($nD$) are possible by taking outer- or Kronecker-products of constituent one-dimensional ($1D$) sequences.

The central slice theorem \cite{KakSlaneyBook} ensures that the Fourier transform of the projected sums of an $nD$ array, as well as the autocorrelation and moment properties, are preserved by projection to $(n-1)D$. The slice theorem ensures the aperiodic autocorrelation of the $(n-1)D$ projected view of an $nD$ array is identical to the $(n-1)D$ projected view of the $nD$ aperiodic autocorrelation. If an $nD$ array has a delta-like autocorrelation, then the projected array in $j$ dimensions also has delta-like autocorrelation, for all $n < j \le 1$.
\subsection{Projection ghosts}
A ghost (or switching element or phantom) in discrete tomography \cite{guedon2009mojette} is an array of pixels with signed values arranged on a regular grid so that all parallel projected rays sum to zero along each of $M$ given discrete directions, $p_i:q_i$, for $1 \le i \le M$. Having zero projections means that (arbitrarily scaled) copies of a ghost can be added anywhere inside any array of data or image values (of same size or larger) without changing the content of those $M$ views. It is relatively easy to construct a ghost in $M$ directions (for any positive integer $M$) as a sum of vector-shifted primitive ghosts (starting with +1,-1) under alternating signs \cite{guedon2009mojette}.  Zero-sum projection ghosts can be trivially modified to produce a sharp delta-like projection by adding a single `bright' pixel at the center of the ghost, but this would circumvent the usual requirement (particularly for `watermarking') for the arrays to be embedded into data with minimal visibility. The sharp projected intensity could be made less visible by distributing the same peak intensity along a single discrete line, parallel to the central ray direction, but then the array would not preserve the ghost property under projection in the other ghost directions. Nor would the array autocorrelation remain delta-like.
\subsection{Huffman arrays exhibit a delta-like projection}
The Mojette transform, $M_{p:q}(t)$, is used to perform the aperiodic discrete projection \cite{guedon2009mojette} of an array or image $I(x, y)$ at angle $p:q$. The integers $t$ label each projected ray. The two-dimensional ($2D$) Huffman array, $H_7\star H_7$, when Mojette-projected at the discrete angle $-1:1$ (-45 degrees) produces a single delta-like response (that matches the $1D$ autocorrelation of $H_7$): $M_{-1:1} = [-1, 0, 0, 0, 0, 0, 18, 0, 0, 0, 0, 0, -1]$. However, a $2D$ array embedded within a larger block of data needs to have two or more delta-like projections to fix the array position through the intersection of those back-projected central rays.
\subsection{Huffman array autocorrelations and spectral flatness}
The Fourier transform $\mathcal{F}$ of a discrete delta function, by definition, is constant over all Nyquist frequencies. If the autocorrelation $C$ of $S$ is delta-like, then $C$ itself is spectrally flat\footnote{Spectral flatness, defined as $(\textrm{max}(F)-\textrm{min}(F))/\textrm{mean}(F)$ for $F = |\mathcal{F}(H_N)|$, gets larger linearly with $n$ for $nD$ arrays built from $1D$ Huffman sequence $H_N$ using the outer-product or by higher order correlations. Smaller values means flatter here.}, and the sequence $S$ must also be flat, as, by the convolution theorem, $|\mathcal{F}(C)| = |\mathcal{F}(S)| \times |\mathcal{F}(S)|$.

The implied inverse of the above statement is true only when the sequence $S$ is spectrally flat under aperiodic boundary conditions. A sequence $S$ that is perfectly flat under periodic conditions \cite{Blake2DPerfectArrayPaper} (for example the uniform amplitude complex Chu sequences \cite{Chu1972}), relies heavily on the periodic (wrap-around) cyclic symmetry to achieve spectral flatness. When such sequences are applied under aperiodic boundary conditions, the loss of assumed periodicity worsens significantly the spectral flatness, as well as lowering the correlation merit factor and peak-to-sidelobe ratio. Huffman sequences retain their `perfect' correlation and spectral flatness properties under both periodic and (zero-padded) aperiodic boundary conditions.

We have seen \cite{SvalbeTCI2020} that arrays made from the $2D$ outer-product $H_{N\times N}$ of Huffman sequence $H_N$ produce a delta-like projection in one direction (for example $-1:1$). Taking the $3D$ outer-product of $H_N$ continues this theme. Projecting the array $H_{N\times N \times N}$ along the $x$, $y$, and $z$ axes produces $2D$ arrays that have a single delta-like projection at angles $1:1$ or $-1:1$.  Projecting $H_{N \times N \times N}$ along the main diagonal (angle $1:1:1$) produces a $2D$ array that has a single delta-like projection at angle $1:0$.

We next show how to build $nD$ Huffman-like arrays that, when projected at discrete angles $p:q:r:...$, for integers $p, q$ and $r$, produce a $2D$ array that has delta-like $1D$ profiles when projected at three discrete angles. We use higher order correlations of Huffman sequences to produce those $nD$ arrays.
\section{HIGHER ORDER CORRELATIONS}
The higher order autocorrelation, $C$, of a sequence $S$ is defined as the sum of shifted products for multiple and orthogonal relative integer shifts $\{t, u, v\cdots\}$, where the summation is over each shift index and the shifts run from $-L$ to $L$ for sequence length $L$:
\begin{align}\label{AcorrHighOrder}
C(t,u,v,\cdots)=\sum{S(i)S(i+t)S(i+u)S(i+v)\cdots.}
\end{align}
For a triple correlation of $S$ with length $L$, $C(t, u)$ is a $2D$ array of size $(2L-1)\times(2L-1)$; the Fourier transform of a triple correlation is often called the bispectrum. As the Huffman arrays have a near-to-flat spectrum, the summed products for each index inside the higher order correlation will be jointly and severally flat. This makes $C$ spectrally flat through copies of the $1D$ correlations that emerge along each orthogonal shift dimension.	
	
\begin{figure}[htb!]
\centering
\includegraphics[width=0.9\columnwidth]{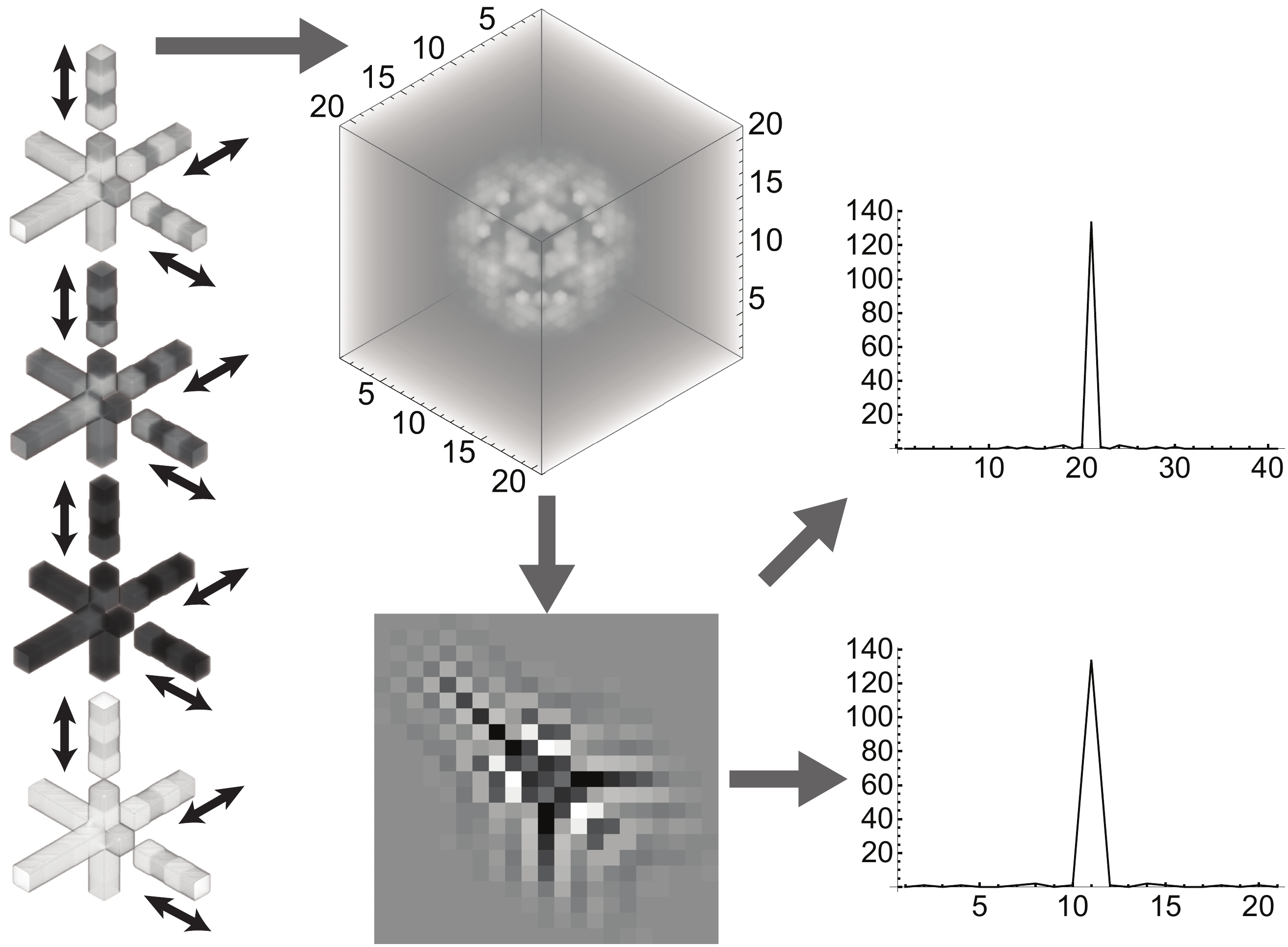} \caption{Delta-like projections arising from a spectrally flat high order correlation function.  To the left of the schematic, a fourth order correlation of an 11 element sequence $H_{11}$ is depicted as a weighted sum of 21 triply-shifted products (4 shown), resulting in the upper $3D$ distribution viewed at angle $1:1:1$.  Summation of the quad-correlation into the page yields the $2D$ pixelated array below, which contains $1D$ delta-like projections over three directions (two shown). Flatness of the Fourier spectrum is preserved throughout.}
\label{fig:Fig2}
\end{figure}

When used to localize an array embedded in a larger block of data by the intersection of back-projected delta-like views, the delta-like angles should have widely spaced, preferably orthogonal directions. Construction of a compact and practical $2D$ array that has delta-like projected views at angles 0, 45 and 90 degrees is portrayed in Fig.~\ref{fig:Fig2}.	

\begin{figure}[htb!]
\centering
\includegraphics[width=1.0\columnwidth]{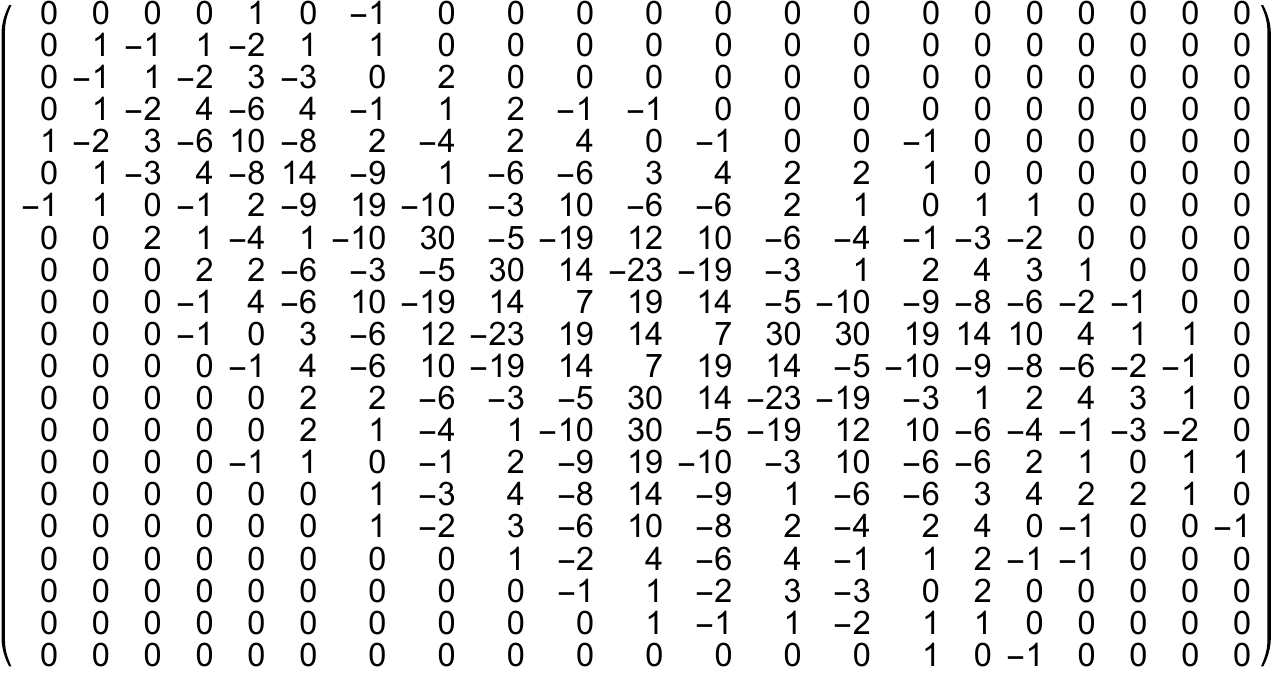}\caption{The $3D$ quad-correlation $C_{11}(t,u,v)$ computed from Eq.~\ref{AcorrHighOrder} as $\sum{H_{11}(i)H_{11}(i+t)H_{11}(i+u)H_{11}(i+v)}$ and projected in the $2D$ plane at the angle $1:1:1$.  The scaled and rounded numerical values correspond to the gray levels of pixelated intensities in Fig.~\ref{fig:Fig2}.}
\label{fig:Fig3}
\end{figure}

The pixelated image within Fig.~\ref{fig:Fig2} shows the $2D$ array obtained from the quad correlation of the Huffman sequence $H_{11} = [1, 2, 2, 4, 6, -1,-6, 4, -2, 2, -1]$ after projection of the $3D$ quad-correlation array at angle $1:1:1$. To make the array less visible, the projected correlation values (that initially ranged from -2530 to +3366) were down-scaled (by 111.5) and rounded to span the integer range -23 to +30.  The resulting $21\times21$ 6-bit array forms identical delta-like projections at the three angles $\{0:1, 1:1, 1:0\}$, where the central ray peak value is 134 and the off-peak values are mostly 0, rarely 1 or 2. The final $2D$ array has aperiodic autocorrelation peak-to-sidelobe ratio of 203 and a merit factor of 424, with Fourier spectral flatness of 0.16.  The integer values of this $2D$ array are shown in Fig.~\ref{fig:Fig3}.

\begin{figure}[htb]
\centering
\includegraphics[width=1.0\columnwidth]{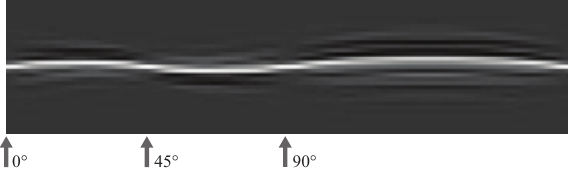} \caption{The Radon transform, over 0 to 179 degrees in 1 degree steps, of the array in Fig.~\ref{fig:Fig3} is delta-like at 0, 45 and 90 degrees and remains close to delta-like in each projected view. }
\label{fig:Fig4}
\end{figure}

Fig.~\ref{fig:Fig4} is the `continuum' Radon transform (projected views in 1 degree steps from 0 to 179 degrees) of the array in Fig.~\ref{fig:Fig3}. It shows sharp delta-like peaks at 0, 45 and 90 degrees. It exhibits less sharp but still delta-like profiles at all view angles, with each profile peak located close to the central ray.
These arrays are robust to affine transforms, for example rotation of the array in Fig.~\ref{fig:Fig3} by $1:1$ produces a $21\times35$ array that has identical correlation metrics but delta-projects at -45, 0 and +45 degrees. 

\begin{figure}[htb]
\includegraphics[ width=1.0\columnwidth]{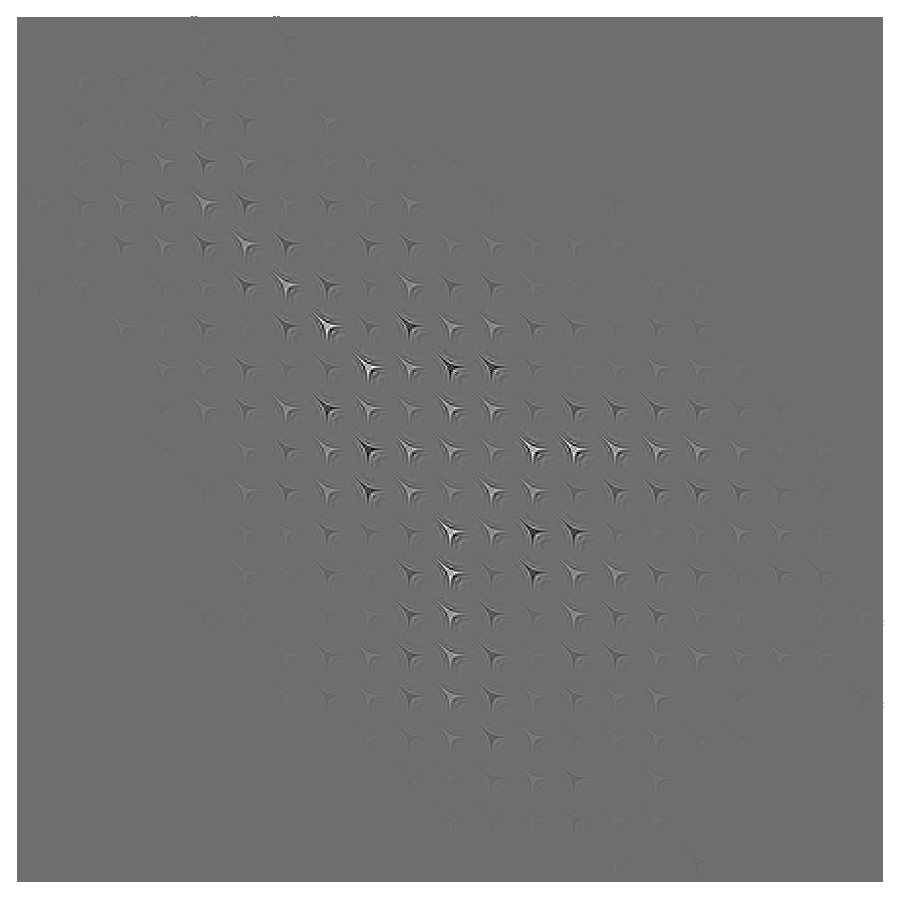}\caption{This $441 \times 441$ $2D$ array inherits the correlation metrics from the single $21\times21$ basis array and is delta-like at the same projection angles $\{0:1, 1:0, 1:1\}$. The array is comprised of integer elements ranging from -690 to 900. 
}
\label{fig:Fig5}
\end{figure}

A $2D$ outer-product projects to its constituent $1D$ arrays for the column or row sums, up to a scale. Treating the delta-like arrays as an approximate Kronecker-basis allows the design of $2D$ arrays with this property over specified directions, not necessarily orthogonal. As a novel extension of this work, one could encode two desired projections (albeit with equal sums) by inverting a system of linear equations to solve for the superposed placements of $2D$ delta-like arrays, each delta-like along one of two chosen views.  Fig.~\ref{fig:Fig5} shows a basic example of this idea, where $2D$ delta-like arrays are weighted and superposed to synthesize a larger matrix that is delta-like along the same directions as those of the constituent basis.  While the method can encode an arbitrary pair of projections when solving for shifts of superposed basis arrays, the particular demonstration in Fig.~\ref{fig:Fig5} avoids overlap of the constituent arrays by forming the $2D$ Kronecker product of the matrix in Fig.~\ref{fig:Fig3}. 
\section{SUMMARY AND CONCLUSIONS} \label{sec: Conclusion}
This Letter demonstrates a general method, based on higher order correlations, to construct integer valued arrays that mimic the behavior of a discrete delta function under both projection and autocorrelation. Here three projected views of the array are delta-like, i.e. mostly zero except at the central ray. The delta-like behavior of these diffuse arrays is evident in both discrete and `analogue' projected views. 

We have found by numerical experiment that the angles over which these delta-like views are formed can be very flexible, as they depend only on the integer values selected for the discrete angles $\{p:q:r:\cdots\}$ used to reduce an $nD$ array to $2D$. Further work is needed to establish if more than three simultaneous delta-like projections can be formed from any one diffuse array. Empirically, we found some simple links to find these angles, which typically include $\{1:1, q:r, (pr+q):(qr), (q+r):p\}$.

The height of the centrally projected ray is determined by the sum of the correlation values along that ray. It may be possible to build larger but flatter Huffman-like arrays, (`quasi-Huffmans') where the central ray of the projected higher order correlation remains large whilst the array values themselves are more uniform and lower, to keep the embedded array values diffuse and less visibly intrusive relative to the surrounding data.
As these arrays are built from the higher order correlations of Huffman (or Huffman-like) sequences, even after projection, they retain their optimal aperiodic correlation properties and close-to-spectrally flat Fourier magnitudes. 

Where these arrays are used for image encoding or watermarking, the position of the embedded array can be estimated from simple back-projection (provided the image background does not dominate the image sums) and confirmed using the more robust but computationally complex decoding via decorrelation (particularly when these delta-like arrays are embedded within large data sets).
\section*{Acknowledgment}
\addcontentsline{toc}{section}{Acknowledgment}
Preliminary results were shown at the Meeting on Tomography and Applications, Discrete Tomography, Neuroscience and Image Reconstruction - 15th Edition, May 4-5, 2021.  Authors IDS and TCP are grateful for feedback during that (online) Milan conference. 
\bibliographystyle{IEEEtran}

\bibliography{refs}

\end{document}